\documentstyle[aps]{revtex}

\begin{document}

\title{A general approach to the localization of unstable periodic orbits
in chaotic dynamical systems}
\author{P. Schmelcher} 
\address{Theoretische Chemie, Physikalisch-Chemisches Institut, Universit\"at 
Heidelberg,\\
Im Neuenheimer Feld 253, 69120 Heidelberg, Germany}
\author{F. K. Diakonos}
\address{Department of Physics, University of Athens\\
GR-15771 Athens, Greece}

\date{\today}
\maketitle

\begin{abstract}
We present a method to detect the unstable periodic orbits of a multidimensional
chaotic dynamical system.
Our approach allows us to locate in an efficient way the unstable cycles of, in principle,
arbitrary length with a high accuracy. Based on a universal set of linear 
transformations the originally unstable periodic orbits are transformed into stable
ones and can consequently be detected and analyzed easily. This method is applicable
to dynamical systems of any dimension and requires no preknowledges with respect to the
solutions of the original chaotic system. As an example of application of our method
we investigate the Ikeda attractor in some detail.
\end{abstract}

\pacs{..Dr, ..+i}
\section{Introduction}
The finding that the unstable periodic orbits provide a skeleton for the
organization of the very complex chaotic dynamics can be considered as one
of the major advances in our understanding of the behaviour of nonlinear
dynamical systems during the past ten years. Many physical quantities
of these systems can be determined by knowing the positions/properties of the
unstable periodic orbits living in the chaotic sea. For strange attractors belonging to
dissipative systems the quantitative description of their structural
properties in terms of unstable periodic orbits has been successfully used for
several dynamical systems \cite{Cvi1,Cvi2,Gre1,Ott1,Lat1}. 
This reflects the importance of the cycles for the analysis
and decoding of the dynamics on the attractor. On the other hand knowing the
positions of the cycles in phase space one can use them to control 
chaotic dynamical systems \cite{Ott2}. Moreover the quantum mechanical
properties of classically chaotic conservative systems possess in the
semiclassical regime a series expansion with respect to the lengths and
the stability coefficients of the periodic orbits (see ref.\cite{Gut1}
and references therein).
Since chaotic behaviour is an intrinsic property of many
dynamical systems periodic orbit theory possesses numerous applications
in different areas of physics.

Assuming that the cycles represent the skeleton for the chaotic dynamical evolution of
dissipative systems one 
can use an appropriate series expansion in terms of their period to determine 
observables like fractal dimensions, average Lyapunov exponents, entropies or the
invariant measure of the corresponding strange attractor \cite{Cvi1,Cvi2,Gre1,Ott3,Eck1}.
The convergence properties of this 
expansion allow us to test the applicability of current periodic orbit 
theory for the properties of dissipative dynamical systems.
The most important and at the same time most difficult step 
of this procedure is to find the location of the unstable cycles in the chaotic sea:
in spite of the fact that their number is growing exponentially with increasing period
and that they are dense in the chaotic sea the unstable cycles are of measure zero whereas the chaotic 
orbits are (for a fully chaotic system) of measure one. However, once the location of the orbits 
is known it is straightforward to extract their properties from the underlying equations of motion.
It is therefore not surprising that a number of recent works deal with the development of efficient
methods and strategies for the detection of the periodic orbits \cite{Bih1,Gra1,Han1}. 

The application of the Newton-Raphson method to find the unstable
cycles requires a good guess for the starting point of the iterative procedure.
For cycles of higher period 
this is only possible if one uses a very fine grid of initial conditions on the attractor. 
Consequently this approach involves a huge numerical effort for more than two dimensions and 
is therefore in general not feasible. Alternative methods can be found 
in the literature but their applicability is limited to special low-dimensional systems
\cite{Bih1,Han1,Cvi3}.   

The purpose of the present investigation is to establish a generally 
applicable as well as reliable and accurate method in order to determine the unstable periodic orbits in a
chaotic  dynamical system \cite{Schm1}. 
The basic idea of our approach is to use an universal set of linear
transformations in order to transform the unstable periodic orbits of the original system
into stable periodic orbits which occupy the same positions in coordinate space.
The stable periodic orbits of the transformed systems can then be found by simply iterating
these dissipative transformed systems. 
It can be shown that such transformations exist in general and
possess very restrictive and simple forms as well as an appealing geometrical
interpretation in terms of a vector field which is organized through the
positions of the cycles.  
To demonstrate the efficiency of our method we apply it to the attractor of the two dimensional Ikeda map
\cite{Ike} which describes the dynamics of an optical field in a ring cavity.
We locate the unstable periodic orbits up to period $13$ of the 
Ikeda strange attractor. Subsequently we derive their stability
coefficients and use the expansions according to periodic orbit 
theory in order to determine the average Lyapunov exponents, the fractal dimension as well as the 
topological entropy of the Ikeda attractor. 

The paper is organized as follows :
In section 2 we present the general theoretical framework of our method and 
discuss in more detail the case of a two dimensional system. Futhermore
we present a universal algorithm for determining the unstable
periodic orbits of a multidimensional chaotic dynamical system.
In section 3 we use this algorithm to locate the unstable
periodic orbits of the Ikeda attractor. From their location we derive the
corresponding stability coefficients which can be used to determine the 
average Lyapunov exponents, the fractal dimension and the topological entropy of the 
attractor. 
In section 4 we summarize the main aspects of our approach and briefly report
on possible applications as well as perspectives related to the general problem
of solving nonlinear equations.

\section{The theoretical framework}

Let us consider a $N$-dimensional discrete fully chaotic dynamical system 
given by:
\begin{equation}
U:~~~~~~~~~\vec{r}_{i+1}=\vec{f}(\vec{r}_i)
\label{eq:dynsysu}
\end{equation}
$U$, being fully chaotic, possesses only unstable fixed points (FP).
Since points of a periodic orbit of period $p$ are FP of the
$p-th$ iterate $\vec{f}^{(p)}$
the term FP stands in the following for periodic orbits in
general, i.e. for orbits of any period. We thereby have to replace $\vec{f}$
in eq.(1) by $\vec{f}^{(p)}$. 
Our goal is to construct from the map $U$ in eq.(\ref{eq:dynsysu})
different dynamical systems $\left\{ S_k |k=1,...,M \right\}$ (see below)
which possess FP at the same positions as $U$ but instead
of being unstable they have become stable in the dynamical systems  $\left\{S_k\right\}$.
The corresponding transformations $\left\{ L_k: U \rightarrow S_k \right\}$ are
required to preserve the number of FP, i.e. no additional orbits
should be created.
The transformations $\left \{ L_k \right \}$ therefore change the stability properties but not
the location of the FP. The index $k$ stands for the 
possibility that different orbits may require different transformations 
for their stabilization.
However, as we shall see below, each type of transformation will stabilize not
only a single FP but a whole rather general class of infinitely many FP
ranging up to arbitrarily high periods. 
If we succeed with our plan then the search for the positions of the FP
of the system $U$ becomes a manageable task: because 
of the stability of the FP in the transformed dissipative systems $\left\{S_k\right\}$ each
trajectory of $S_k$ after some iterations closely approaches a FP 
$\vec{r}_F$. Per construction $\vec{r}_F$ is then also a FP of the 
system $U$ and we therefore know the position of the FP in the
original system $U$. 

To fulfill the requirement of the one to one correspondence between
the FP of $U$ and $S_k$ the transformation $L_k$
should in general be linear. Consequently $S_k$ takes on the
following appearance:
\begin{equation}
S_k:~~~~~~~~~\vec{r}_{i+1}=\vec{r}_i+\bbox{\Lambda}_k (\vec{f}(\vec{r}_i)-
\vec{r}_i)
\label{eq:dynsyss}
\end{equation}
where $\bbox{\Lambda}_k$ is a constant invertible $N \times N$ matrix. 
The definition (\ref{eq:dynsyss}) satisfies the one to one
correspondence of the FP of $U$ and those of $S_k$: If $\vec{r}_i=\vec{r}_F$ is
a FP of $U$ then the parenthesis on the right hand side 
of eq.(\ref{eq:dynsyss}) vanish and therefore 
$\vec{r}_F$ is also a FP of $S_k$. On the other hand
if $\vec{r}_F$ is a FP of $S_k$ and since $\bbox{\Lambda}_k$ is nonsingular
the parenthesis on the right hand side of eq.(\ref{eq:dynsyss}) must be equal to zero for 
$\vec{r}_i=\vec{r}_F$ which implies that $\vec{r}_F$ is also a FP 
of $U$. Thus the dynamical laws $U$ and $S_k$ possess FP at 
identical positions in space.

Our next step is to stabilize the FP of the transformed systems 
$S_k$ by suitable choices for $\bbox{\Lambda}_k$.
Different unstable FP of the map $U$ are then, in general, stable
in different transformed dynamical laws $S_k$. However, as we shall see below,
the number of matrices $\left\{\bbox{\Lambda}_k \right\}$ necessary to achieve
stabilization of {\it {all}} FP is very small.
It turns out
that if the absolute values of the elements of the matrices 
$\bbox{\Lambda}_k$ are 
sufficiently small ($\vert \lambda^{(k)}_{ij} \vert \ll 1,~i,j=1,..,N$) 
then there exists
a universal set of very restrictive matrices such that at 
least one matrix $\bbox{\Lambda}_k$ belonging to this set transforms (via
eq.(\ref{eq:dynsyss})) a given unstable FP of $U$ to a stable FP 
of the corresponding map $S_k$. In order to determine this set of matrices 
let us consider the 
stability matrices $\bbox{T}_U$ and $\bbox{T}_{S_k}$ of $U$ and $S_k$ which 
obey the following relation:
\begin{equation}  
M_k:~~~~~~~~~\bbox{T}_{S_k}=\bbox{1} + \bbox{\Lambda}_k (\bbox{T}_U- \bbox{1})
\label{eq:stama}
\end{equation}
For $\bbox{T}_U$ we assume that it is real, invertible and
diagonalizable. 
Since $\vec{r}_F$ is an unstable FP at least one of the
eigenvalues of $\bbox{T}_U$ at $\vec{r}_F$ must possess an absolute value 
greater than one. In order
to stabilize $\vec{r}_F$ we proceed in two steps: first we use the
parametrization $(\bbox{\Lambda}_k)_{ij}=(\lambda \bbox{C}_k)_{ij}$ with 
$1 \gg \lambda > 0$
and $C_{ij}=O(1)$. The matrix $C_{ij}$ has to be chosen such that the
real parts of all eigenvalues of the matrix 
$\bbox{C}_k \cdot (\bbox{T}_U-\bbox{1})$ are negative. If this is 
achieved then 
the next step is to use a sufficiently small value for the parameter 
$\lambda$, such that the eigenvalues of the matrix 
$\bbox{T}_{S_k}=\bbox{1}+\lambda \bbox{C}_k (\bbox{T}_U-\bbox{1})$ have absolute values
less than one. It can be shown that this is always possible if $\lambda$ is 
sufficiently small: for sufficiently small values of $\lambda$ the leading contribution
with respect to $\lambda$
in the absolute values of the, in general complex, eigenvalues is of order
$\lambda^1$ and emerges from the real part of the eigenvalues of the transformed system.
Consequently we have transformed the unstable FP $\vec{r}_F$ of $U$ via 
$L_k$ into a stable FP of $S_k$ thereby keeping its position fixed.

Next let us derive possible sets of matrices $\{\bbox{C}_k\}$ which, according to the 
above discussion, transform the signs of the real part of the eigenvalues of
$\bbox{T}_{U} -1$ and consequently allow us to stabilize any configuration of unstable FP.
First we provide a rather general set of matrices and subsequently
we will specialize to a much more restrictive and simpler set.
For any given $\bbox{T}_{U}$ it is possible to find an involutory matrix 
$\bbox{C}_k$ defined through $\bbox{C}_k^2=\bbox{1}$ such that 
$\bbox{A}=\bbox{C}_k(\bbox{T}_U- \bbox{1})$  has eigenvalues with negative 
real parts. The proof reads as follows: Assuming the diagonalizability of the 
matrix $\bbox{B}=\bbox{T}_U- \bbox{1}$ there exists a similarity transformation
$\bbox{P}$ such that $\bbox{B_D}=\bbox{P}^{-1} \bbox{B} \bbox{P}$ 
is diagonal. We have then:
$$\bbox{P}^{-1} \bbox{A} \bbox{P} =\bbox{P}^{-1} \bbox{C}_k \bbox{P} \cdot
\bbox{B_D}$$
We can therefore always choose a diagonal matrix $\bbox{C_D}_k=
\bbox{P}^{-1} \bbox{C}_k \bbox{P}$ with elements $\pm 1$ on its diagonal such
that $\bbox{C_D}_k \cdot \bbox{B_D}$ has eigenvalues with negative real part.
Such a matrix $\bbox{C_D}_k$ is according to its definition involutory. 
Due to the invariance of the eigenvalue spectrum of a matrix with respect to 
similarity transformations, the matrix $\bbox{A}$ has then also eigenvalues
with negative real part. In addition if $\bbox{C_D}_k$ is involutory then any
matrix resulting from $\bbox{C_D}_k$ through similarity transformations, and in
particular $\bbox{C}_k$, is also involutory. We can therefore always find
an involutory matrix $\bbox{C}_k$ given by $\bbox{C}_k=\bbox{P} \bbox{C_D}_k
\bbox{P}^{-1}$ such that $\bbox{C}_k(\bbox{T}_U- \bbox{1})$ possesses 
eigenvalues with negative real parts. 

The set of involutory matrices is for our purposes rather general.
For reasons of applicability as well as practical efficiency of our
method it is therefore desirable to find a smaller set
of transformation matrices.
In fact it turns out that a much more restricted form for 
the matrices $\bbox{C}_k$ is sufficient in order to achieve stabilization
of any FP via the transformations $\left\{L_k\right\}$, namely all
the matrices corresponding to special reflections and rotations in space. 
The elements of this new class of matrices $\left \{ \bbox{C}_k \right\}$
are $C_{ij}^{(k)} \in \{0,\pm 1\}$ 
and each row or column contains only one element which is different from zero.
The matrices $\bbox{C}_k$ are therefore orthogonal. 
The number $a_N$ of such matrices in $N$ dimensional space is given by 
$a_N=N!$ $2^N$.

To illuminate our approach we discuss in the following the case 
$N=2$ in more detail. In this case $U$ is given by:
\begin{eqnarray}
x_{i+1}&=&f(x_i,y_i)  \nonumber \\
y_{i+1}&=&g(x_i,y_i) 
\end{eqnarray}
and represents a fully chaotic 2-D map with a
$2 \times 2$ stability matrix $\bbox{T}_U$. Let us denote with $\rho_{1,2}$  
the eigenvalues of $\bbox{T}_U$. Being fully chaotic $U$ possesses only
hyperbolic FP. The choice of the matrix $\bbox{C}_k$ appropriate for
the stabilization of a particular FP depends on the class to which
the hyperbolic FP belongs.
According to the above discussion there are $a_2=8$ possible
matrices $\left\{\bbox{C}_k\right\}$ as candidates to be used for this stabilization process.
With some algebra one can find the matrices which achieve 
the stabilization of the different types of unstable FP:
\begin{itemize}
\item{For hyperbolic FP with reflection and $\rho_1 < -1,~ -1 < \rho_2 < 0$ 
$\Rightarrow$ $\bbox{C}_1=\bbox{1}$}
\item{For hyperbolic FP with reflection and $\rho_1 < -1,~ 0 < \rho_2 < 1$ 
$\Rightarrow$ $\bbox{C}_1=\bbox{1}$}
\item{For hyperbolic FP with reflection and $\rho_1 > 1,~ -1 < \rho_2 < 0$
or hyperbolic FP without reflection ($\rho_1 > 1,~ 0 < \rho_2 < 1$) 
we have to distinguish between the following three cases: \\
(a) $\displaystyle{\frac{\partial f}{\partial x}} \mid_{FP}~>~
\displaystyle{\frac{\partial g}{\partial y}} \mid_{FP}~$ 
$\Rightarrow$ $\bbox{C}_2 =
\left( \begin{array}{cc} -1 & 0 \\ 0 & 1 \end{array} \right)$ \\
(b) $\displaystyle{\frac{\partial f}{\partial x}} \mid_{FP}~<~
\displaystyle{\frac{\partial g}{\partial y}} \mid_{FP}~$ 
$\Rightarrow$ $\bbox{C}_3 =
\left( \begin{array}{cc} 1 & 0 \\ 0 & -1 \end{array} \right)$ \\
(c) $\displaystyle{\frac{\partial f}{\partial x}} \mid_{FP}~=~
\displaystyle{\frac{\partial g}{\partial y}} \mid_{FP}~$ 
$\Rightarrow$ $\bbox{C}_4 =
\left( \begin{array}{cc} 0 & -1 \\ -1 & 0 \end{array} \right)~~$ or 
$~~\bbox{C}_5 =
\left( \begin{array}{cc} 0 & 1 \\ 1 & 0 \end{array} \right)$}
\end{itemize}

With the above choices for $\bbox{C}_k$ and sufficiently small
values of the parameter $\lambda$ we can stabilize any hyperbolic
FP of a chaotic 2-D system. The only exception where our
approach does not work is the case of a parabolic FP, i.e.
for parallel flow ($\rho_1=1 \vee \rho_2=1$). 
However this case is of no interest to the present investigation.

It is a major advantage of the present approach that a single
matrix $\left\{\bbox{C}_i\right\}$ is responsible for the stabilization
of an infinite number of unstable FP belonging to periodic orbits
of arbitrarily high periods. For example: the matrix $\bbox{C}_1$
stabilizes hyperbolic FP (periodic orbits) with reflection with
respect to both invariant manifolds, independent of their
period. We remark that only 5 of the 8 matrices $\bbox{C}_k$ 
are necessary to stabilize all hyperbolic FP in two dimensions.
The matrices $\bbox{C}_4$ and $\bbox{C}_5$ typically
stabilize no or only a very few additional FP which is due to the
fact that the condition $\displaystyle{\frac{\partial f}{\partial x}} \mid_{FP}~=~
\displaystyle{\frac{\partial g}{\partial y}} \mid_{FP}~$
is met only very rarely. 
A similar statement holds also for higher dimensional systems:
only a subset of the above given class of $a_N$ matrices is actually
necessary in order to achieve the stabilization of any given FP.
According to the above we have shown a one way criterium with respect
to the stabilization of the FP. For example:
any hyperbolic FP with reflection ($\rho_1 < -1, -1 < \rho_2 <0$) in two dimensions can
in particular be stabilized
by using the matrix $\bbox{C}_1$ but might also be stabilized by using other
matrices of the set $\bbox{C}_k$.
The explicit statements saying which matrices stabilize which types of FP become increasingly
more complex with increasing dimension of the dynamical system under investigation.

Having presented the basic features of our method we discuss now briefly 
how it can be used to detect the unstable FP of a given 
$N$-dimensional discrete fully chaotic dynamical system. 
Using eq.(\ref{eq:dynsyss}) we transform the given dynamical system $U$ into a new 
system $S_k$. For the matrix $\bbox{\Lambda}_k=\lambda \bbox{C}_k$ we use a 
sufficiently small value of $\lambda$ and $\bbox{C}_1$ from 
$\{ \bbox{C}_k \},~k=1,..,a_N$. 
Iterating an arbitrary initial point $\vec{r}_0$ with $S_1$ from 
eq.(\ref{eq:dynsyss}) forward in time we observe the following characteristic
behaviour: either the trajectory runs to a stabilized FP $\vec{r}_F$ with
steps of continuously decreasing size or, if stabilization is not achieved 
using $\bbox{C}_1$, the trajectory chaotically evolves on the attractor or
escapes to infinity. 
In order to get all the FP of $U$ this procedure has to be
repeated for a representative set of starting points which covers in a crude way (see below)
the phase space of the system. Subsequently we perform the same procedure for the next matrix 
$\bbox{C}_2$ of the set $\bbox{C}_k$, etc... until all the matrices 
$\{\bbox{C}_k\}$ have been used. Following this procedure we obtain the complete set of FP 
of $U$.
In order to find the unstable FP of period $i$
we simply have to replace $\vec{f}$ in eq.(\ref{eq:dynsyss}) by its i-th 
iterate $\vec{f}^{i}$. One must however be careful with respect to the choice of the 
value of $\lambda$ in $\bbox{\Lambda}_k$ because it is indirectly related to 
the stability coefficients of the desired cycle. 
We therefore have to use decreasing values of $\lambda$
with increasing periods of the cycles we would like to detect.
However later on in the present section we will provide a continuous formulation of the
transformed systems which is independent of the value of the parameter $\lambda$.
Due to the fact that different
kinds of FP (e.g. hyperbolic FP with and different kinds without reflection)
are stabilized by different matrices $\bbox{C}_k$ our
stabilization procedure offers the possibility to distinguish
between the different types of FP. 

An important advantage of our stabilization method is its global
character. Even points lying far from the linear neighbourhood of the
stabilized FP are attracted to it after a few iterations of the
transformed dynamical law. 
To each FP of a certain period we can assign its basin of attraction consisting of the
set of starting points which converge towards the stabilized
FP if we iterate them with the transformed dynamical system $S_k$.
With increasing period the typical volume of a basin of attraction 
becomes increasingly smaller and the basins form a geometrically very complex and 
interwoven network which covers the attractor.

Using the above approach to detect the unstable periodic orbits of a given
chaotic dynamical system it turns out that
the number of starting points needed to obtain the periodic orbits of a given
period on the (closure of the) attractor is only a few times more than the
number of cycle points themselves. Suitable starting points for the application
of the transformed dynamical law $S_k$ can be obtained by using, for example,
a chaotic trajectory on the attractor itself or more refined methods which
cover the attractor in a more systematic manner. 
We used a simple empirical strategy in order to seek for the periodic
orbits of a given period. Assuming that a certain number of periodic
orbits of period $p$ have been found we iterate a multiple of the starting
points used in the previous run. 
If no additional periodic orbits show up we take the set of detected
orbits as complete. We emphasize that this is by no means a proof for 
the completeness of the set of detected periodic orbits.
However it turns out that for dynamical systems for which the periodic
orbits are known, like for example the Henon map, the above procedure
yields all periodic orbits.
Because of the fact that each FP possesses a finite basin of attraction (see below and Fig.2)
the above method turns out to be rather insensitive with respect to the specific choice of the set of
starting points covering the attractor in a crude and large-meshed manner.

The transformed dynamical law ($\ref{eq:dynsyss}$) possesses an appealing 
geometrical interpretation.  Let $\{ \vec{r}_j,~j=1,..,p\}$ be a trajectory 
of the original system $U$. At each point of the trajectory we define a vector field 
$\vec{V}_U(\vec{r}_j)=\vec{r}_{j+1}-\vec{r}_j$. The corresponding
transformation $L_k$ represents then a special reflection/rotation of each vector 
$\vec{V}_U(\vec{r}_j)$ combined with a subsequent scale transformation of its 
length with the factor $\lambda$.
Using eqs.(1,2) we obtain for the vector field $\vec{V}_{S_k}$ of the transformed map
\begin{equation}
\vec{V}_{S_k} (\vec{r}) = \bbox{\Lambda}_k \vec{V}_{U} (\vec{r})
\end{equation}
The main feature of the new vector field $\vec{V}_{S_k}$ is its global organization
around those FP which have been stabilized. The flow of the vector
field $\vec{V}_{S_k}$ is centered and organized around
the positions of the stable FP which represent sinks/sources of this vector field.
To illustrate these properties we show
in Fig.1 selected vector fields belonging to a chaotic trajectory on 
the attractor of the Ikeda map (see also section 3).
In Fig.1(a) the vector field $\vec{V}_{S_k}$ 
for the first iterate $\vec{f}^{1}$ is illustrated. 
We use a trajectory of 200 points
and the stabilization matrix $\bbox{\Lambda}_k=0.1 \cdot \bbox{C_1}$.
The global organization of the flow towards the FP 
$\vec{r}_F=(0.5328,0.2469)$ whose position is indicated in Fig.1 by a cross is 
evident. Obviously this property is not restricted or specific for the linear 
neighbourhood of the FP but represents a global feature of the 
dynamical system.
Starting with any point on the attractor the trajectory of the transformed
map moves immediately towards the FP and yields with increasing number of
iterations increasingly accurate values for its position.
In Fig.1(b,c) we show the vector field $\vec{V}_{S_k}$ in the vicinity of
the FP $\vec{r}_{F,1}=(0.5098,-0.6084)$ and 
$\vec{r}_{F,2}=(0.6216,0.6059)$
for the second iterate $\vec{f}^{(2)}$ of the Ikeda map which represent a 
period two cycle of the map. 
In Fig.1(b) 200 points in the neighbourhood of $\vec{r}_{F,1}$ are plotted.
The stabilization matrix $\bbox{\Lambda}_k=0.02~\bbox{C_1}$ was used. In Fig.1(c) we use
the same number of points as well as the same stabilization matrix 
$\bbox{\Lambda}_k$ to illustrate the organization of the vector field around 
$\vec{r}_{F,2}$. Again the positions of the FP are indicated by crosses.
From Figs.1(b,c) it is evident that the flow of the vector field
$\vec{V}_{S_k}$ can exhibit sharp, but still smooth, turns in the immediate
neighbourhood of the FP. The size of these regions of nonlinearity
in general decreases with increasing period of the cycle considered.
  
The fact that the stabilized FP represent the centers of the flow 
of the vector field $\vec{V}_{S_k}$ possesses a counterpart in the original 
chaotic system $U$ which we shall discuss briefly in the following.
If a chaotic trajectory approaches a FP or more precisely enters
the linear neighbourhood of the FP it can be shown that the
maximum of the deflection of the trajectory in the linear regime occurs for
the position closest to the FP. The deflection of the incident
trajectory can be considered as a turning of the trajectory and the point
of maximum deflection (or curvature for continuous systems) can be
considered as a generalization
of the notion of a turning point in one dimension (see refs.\cite{Dia1}).
Different chaotic trajectories approaching the FP from different
directions in space are deflected to different directions of their outgoing
manifold. The common feature of chaotic trajectories in the neighbourhood of a 
FP is the fact that the above-mentioned deflection or turning
occurs with respect to any direction of space.
If the chaotic system $U$ is transformed via the corresponding stabilizing
transformation $L_k$, the resulting dynamical system $S_k$ possesses centers 
of the corresponding flow of the transformed system at the positions of the FP.
$L_k$ can therefore be considered as a transformation from spreading to focusing
flow. 

The above-described method of detecting the unstable FP (periodic orbits) for a given
chaotic dynamical system involves the parameter $\lambda$ which has to be chosen
sufficiently small in order to transform the unstable FP of the
original chaotic system to a stable one via the transformations
${L_k}$. With increasing period of the FP to be located, the parameter $\lambda$ has to
be chosen increasingly smaller in order to achieve stabilization. However we should not choose 
$\lambda$ too small since the convergence of the iterated
transformed dynamical laws (see eq.(2)) to the FP will then become very slow and
involves a waste of computer time.   
There is a simple way to avoid this tuning of the parameter $\lambda$
which makes our approach independent of the parameter $\lambda$.
Taking the limit $\lambda \rightarrow 0$ in eq.(2) we obtain
\begin{equation}
\lim_{\lambda\to 0} \frac{\left(\vec{r}_{i+1}-\vec{r}_{i}\right)}{\lambda} = \dot{\vec{r}}
= \bbox{C}_k (\vec{f}(\vec{r}) - \vec{r})
\end{equation}
This equation represents the continuous formulation of our transformed discrete dynamical systems
in eq.(2). Its solutions possess the same nice properties as those of eq.(2) in the sense that
any stabilized trajectory is running to a stable FP. The important difference however is 
that the solutions of eq.(6) do not depend on the parameter $\lambda$ and it can therefore easily be solved by
using any suitable integration scheme like, for example, a Predictor-Corrector integrator. 

This means in particular that the basin of attraction for a certain FP is now independent of
the choice of the parameter $\lambda$ and represents a characteristics of the originally unstable
FP and the transformation $L_k$. In Figure 2 we show a typical ensemble of trajectories which are
solutions to eq.(6) for a given set of starting points for the
case of the stabilized period two orbit of the Ikeda map (see section 3). Part of the trajectories are running
towards one point of the period two orbit and the other part towards the second point of the orbit.
The boarderline of the two basins of attraction represents an unstable one dimensional manifold
which is the only region of instability concerning the coordinate range of the Ikeda attractor.

We emphasize that the above described method for the detection of the unstable periodic orbits
can be applied to any chaotic dynamical system independent of its dimension and individual properties.
The advantages of our approach are obvious: given any chaotic dynamical system with a skeleton of
unstable periodic orbits we can transform these orbits into stable periodic orbits of a different
transformed dissipative dynamical system. A large class of periodic orbits is thereby stabilized
by the same linear transformation. The positions of the periodic orbits can then be obtained simply
by iterating the transformed system or solving the differential equations of its continuous formulation.
In this way the FP can be determined with in principle arbitrary accuracy.

Finally we mention that our method is by no means restricted to discrete dynamical systems.
Periodic orbits in continuous dynamical systems can be detected by using the Poincare map which
is again a discrete map representing the original continuous dynamical system in a chosen hyperspace.
It is hereby not necessary for the Poincare map to be given in an analytical form. The 
transformed dynamical systems can be obtained from eqs.(2,6) for a numerically given Poincare map.
All the quantities given in these equations are defined in the Poincare section which is a
(N-1)-dimensional space and periodic orbits represent a finite number of points in this space
i.e. are of integer period in the Poincare section. It is therefore not necessary to know the 
period of the orbits in the continuous system in order to detect them in the Poincare
section.

\section{Application to chaotic dynamical systems}

We applied our method to several $2D$ iterative maps like, for example, the
Henon, $2D$-logistic and in particular the more complicated Ikeda map.
In the following we will discuss in some detail our procedure and the corresponding
results for the Ikeda attractor. 

To demonstrate the reliability as well as efficiency of our method we use it to calculate the
unstable periodic orbits of the Ikeda attractor. 
The underlying dynamical law is the Ikeda map \cite{Ike} used in nonlinear
optics to describe the response of a 2-level homogeneous absorber in a ring cavity
to a constant incident light wave. The map is given by:
\begin{eqnarray}
U_I:~~~~~~x_{n+1}&=&1+0.9~(x_n \cos w_n - y_n \sin w_n) \nonumber \\
y_{n+1}&=&0.9~(x_n \sin w_n + y_n \cos w_n)~~~~~~~~~~~~
w_n=0.4-\frac{6}{1+x_n^2+y_n^2} 
\label{Ikeda}
\end{eqnarray} 
The Ikeda attractor is fully chaotic and embedded in two dimensional coordinate space.
Our considerations for the $N=2$ case described in section 2
can therefore be applied directly. In detail we proceed as follows:
Using the set of stabilization matrices 
$\bbox{C}_i,i=1,..,5$ given in section 2 we construct according to eqs.(2,6)
the transformed systems which possess the corresponding stable FP.
Starting from an initial point we next produce iteratively a chaotic trajectory on the attractor
which provides a set of starting points for the trajectories of the transformed dynamical
systems which are solutions to eqs.(2,6).
If a certain trajectory converges to a point $\vec{r}_{F}$ within a given accuracy then this point
is recorded as a FP of the Ikeda map. We repeated this procedure
for the higher iterates of the Ikeda map up to $13-th$ iterate.
The number of starting points on the attractor needed to obtain 
the periodic orbits of period 13 is roughly $5000$. This is particularly impressive if we take into
account that some of the period 13 orbits differ only from their fourth significant
digit on and that the number of FP detected for period 13 is $2522$.
We are therefore in the position to resolve a large number of close lying periodic orbits.
Again we emphasize that we have no rigorous proof for the completeness of the detected orbits
but empirical evidence.

The origin of the success of our stabilization transformation is its global character. 
Thus we need only a very coarse grained lattice of initial conditions to 
cover the attractor.
For the discrete formulation of our transformed dynamical systems (see eq.(2))
the values of the parameter $\lambda$ necessary to achieve stabilization vary from 
$10^{-1}$ for orbits of low period to a few times $10^{-5}$ for periodic orbits of period 13. 
Another important
feature of our approach is the high accuracy which is controlled by the
convergence of the corresponding trajectories. Our calculations yield a 
relative accuracy of $10^{-13}$. Having determined the coordinates
of the cycles we use the Ikeda map in eq.(\ref{Ikeda}) to derive the
corresponding stability coefficients. 

The corresponding results are presented in table I.
The first row shows the number of unstable cycles for a given period $p$ and in the second row
the total number of cycle points of order $N$ is given.
Starting with one period 2, two period 3 and three period 4 cycles their number
increases strongly with increasing period and finally for period 12 and 13 we obtain
110 and 194 cycles, respectively. With the help of the obtained
orbits and their stability properties we are in the position to determine characteristic
quantities of the attractor related to its degree of chaoticity or dynamical 
and geometrical complexity. To get an impression of the quality of the
covering (representation) of the Ikeda attractor through the unstable periodic orbits
we illustrate in Fig.3(a) the set of all the FP of the Ikeda map and 
its higher iterates (up to the $13-th$ iterate). For comparison we present in Fig.3(b) a typical
chaotic trajectory on the attractor which possesses the same total number of 
points (5627). Although the overall picture looks very similar a more careful
comparison of Figs.3(a,b) reveals major differences in the local density
of points. If we compare the distribution of the FP shown in Fig.3(a)
with the Ikeda attractor of Fig.3(b) we realize that there
exist regions within the support of the attractor which are not visited
by periodic cycles up to period 13 at all. In addition there exist regions of high density
of FP (see Fig.3(a)) which do not correspond to regions of high density for the 
attractor itself (see Fig.3(b)). 

In order to perform a quantitative analysis of the Ikeda attractor in terms
of the unstable cycles let us first calculate the measure of the 
exponential increase (with increasing period) of their number or in other words the so called
topological entropy defined through \cite{Kat1}:
\begin{equation}
S_T=\lim_{p \rightarrow \infty} \frac{1}{p} \ln n(p)
\end{equation}
where $n(p)$ denotes the number of FP of $U_I^p$. The values of
$S_T$ are presented in table I (third row) and they can be
used as a rough guide for the completeness of the number of periodic orbits found
for a given period.
An average measure of the strangeness of the attractor is given in terms
of the average Lyapunov exponents defined through \cite{Ott3}:
\begin{equation}
h_{e,c}=\lim_{p \rightarrow \infty} \sum_j \frac{1}{\mu_e(\vec{r}_{jp})}
\ln \mu_{e,c}(\vec{r}_{jp})
\end{equation}
where $\vec{r}_{jp}$ denotes the j-th
FP of the p-th iterate of the map $U_I$ and $\mu_e(\vec{r}_{jp})$
($\mu_c(\vec{r}_{jp})$) is the expanding (contracting) eigenvalue of the
stability matrix at this point.
The two exponents correspond to an average expanding rate ($h_e$) and
an average contracting rate ($h_c$), respectively.
The results as a function of period $p$ are presented in the fourth row of
Table I. The strong fluctuations in successive terms of this expansion
can be seen best in Fig.4(a) where the two Lyapunov exponents are shown
as a function of the period p.
To characterize the strange attractor geometrically we follow \cite{Cvi1}
and try to cover the attractor with slabs of length 1 and width $\frac{1}{
\mu_{c}(\vec{r}_{jp})}$. In this case the corresponding Hausdorff dimension 
$D_o$ can be found by solving the equation:
\begin{equation}
\sum_j \mu_{c}(\vec{r}_{jp})^{D_o^{(p)}-1}=1
\end{equation}
The numerical results with increasing period $p$ are given in the fifth and last row of 
table I. The corresponding graphical illustration is shown in Fig.4(b).
Again strong oscillations are present even for higher periods.  

Summarizing our analysis of the Ikeda attractor we observe a satisfactory 
convergence of the topological entropy up to period 13 but strongly oscillating values
for the fractal dimension as well as average Lyapunov exponents.

\section{Conclusions}

The main objective of the present paper is the development of a general method to
detect the unstable periodic orbits in chaotic dynamical systems. 
The central idea is to convert the unstable
fixed points (periodic orbits) to stable ones without changing their location in space.
We have shown that this can be achieved by a set of linear transformations,
i.e. transformation matrices, which allow the stabilization of any configuration
of unstable hyperbolic fixed points for a given dynamical system.
The numerical calculation of the stable fixed points in the transformed dissipative dynamical
systems is done by either a simple iteration procedure or by solving the corresponding
continuous version of the transformed dynamical laws.

The above approach allows a straightforward application to any analytically or numerically
given dynamical system. Apart from its general applicability the advantages of our method
are the following. 
It is of universal character in the sense that no previous knowledges about the topological
or dynamical behaviour, except the presence of chaos, are required. 
Our method is by no means restricted to lower dimensional (1D or 2D) systems
but can in principle yield periodic orbits for dynamical systems of any dimension.
It allows an efficient convergence to highly accurate values and requires at the same time
only a small set of initial conditions to cover the strange attractor:
a coarse grained covering is sufficient to detect the periodic orbits which are 
within our empirical procedure assumed to be complete.
Nevertheless even close lying
periodic orbits of higher periods can be resolved and, therefore, distinguished through the
evolution of the stabilized system. The underlying stabilization transformation possesses
an appealing geometrical interpretation. While the dynamics of the original chaotic
system in the vinicity of the fixed points is characterized by a 'turning' of trajectories
\cite{Dia1}
in any direction of coordinate space the stabilized system belongs to a vector field which
is centered around specified configurations of sinks/sources. The positions of those
sinks/sources are identical with the positions of the fixed points.
We remark that our method can also be applied to cases where the dynamical system
is not fully chaotic but consists of a chaotic sea filled with islands of regularity.

The above method is not restricted to discrete dynamical systems:
periodic orbits in continuous dynamical systems can be detected by using the Poincare map which
is again a discrete map representing the original continuous dynamical system in a certain 
hyperspace.
As a further perspective and application of the
developed scheme we mention the possibility of solving nonlinear equations in general 
through the iteration of a suitably stabilized version of the original equations.
Thereby we translate the problem of finding the roots of a set of nonlinear equations 
$\vec{f} (\vec{x}) = \vec{0}$ to that of finding the fixed points of the dynamical
law $\vec{F} (\vec{x})= \vec{f}(\vec{x}) + \vec{x}$. Applying the above method to
get a stabilized version of $\vec{F}$ we obtain then by simple iteration the
fixed points of $\vec{F}$ which correspond to the roots of $\vec{f}$.

\vspace*{1.0cm}

{\large \bf{Acknowledgments}}

The European Community (F.K.D.) is gratefully acknowledged for 
financial support.

{}

\vspace*{5.0cm}

\begin{center}
\large \bf {Figure Captions}
\end{center}

\begin{figure}
\caption{(a) The vector field $\vec{V}_{S_k}$ belonging to the Ikeda map on its attractor.
(b) The corresponding vector field of the second iterate of the Ikeda map around the (stabilized) fixed point 
(0.5098,-0.6084) belonging to the period two cycle of the Ikeda map.
(c) The vector field around the other fixed point (0.6216,0.6059)
of the period two cycle. The positions of the fixed points are 
indicated by crosses. 200 points have been used for each subfigure.} 
\end{figure}

\begin{figure}
\caption{Lines which represent an ensemble of trajectories of the transformed map in eq.(2)
for the stabilization of the period 2 orbit of the Ikeda attractor. The circles indicate the
positions of the points of the period 2 orbit and the line connecting the crosses indicates
the boarder line between the two basins of attraction.} 
\end{figure}

\begin{figure}
\caption{(a) The set of all the fixed points of the Ikeda map and its higher
iterates up to period 13.
(b) A typical chaotic trajectory on the Ikeda attractor with the same number
of points (5627) as in (a).}
\end{figure}

\begin{figure}
\caption{(a) The average Lyapunov exponents as a function of the period $p$.
(b) The fractal dimension $D_o$ of the Ikeda attractor as a function of the period $p$.
See section 3.}
\end{figure}

\vspace*{1.0cm}

\begin{center}
\large \bf {Table Captions}
\end{center}

{\bf{Table I:}} The number of cycles with period N, the total number of cycle points of order N,
the topological entropy, the Lyapunov exponents as well as the fractal dimension for 
N=1,...,13 (see section 3). 

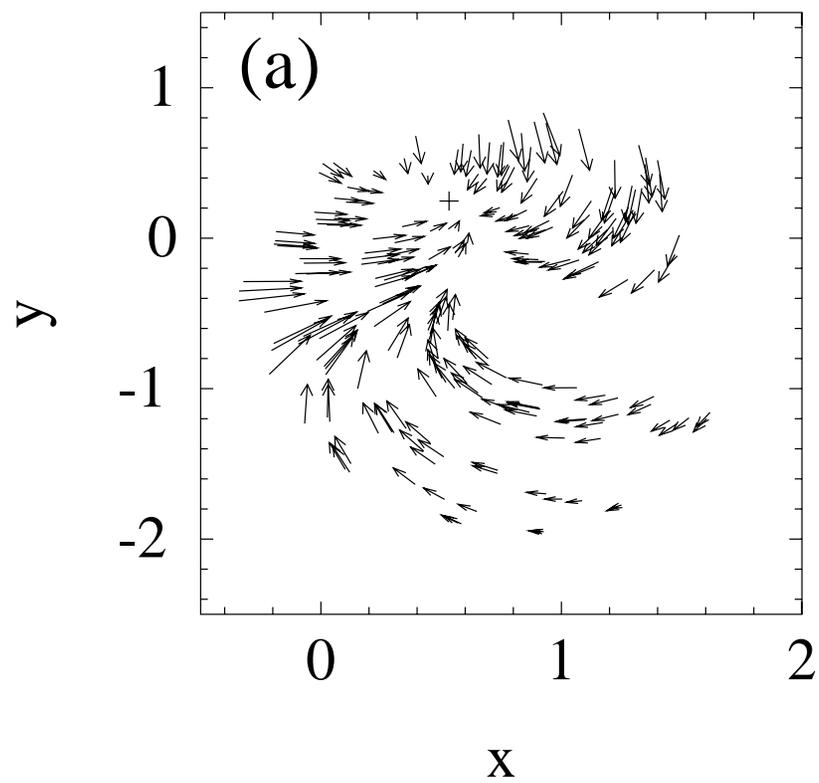
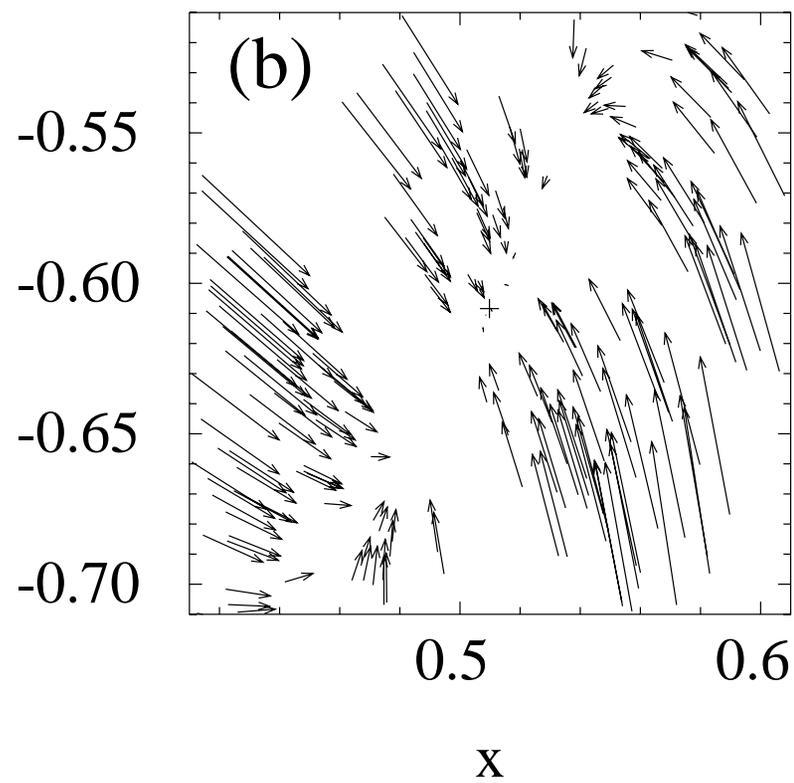
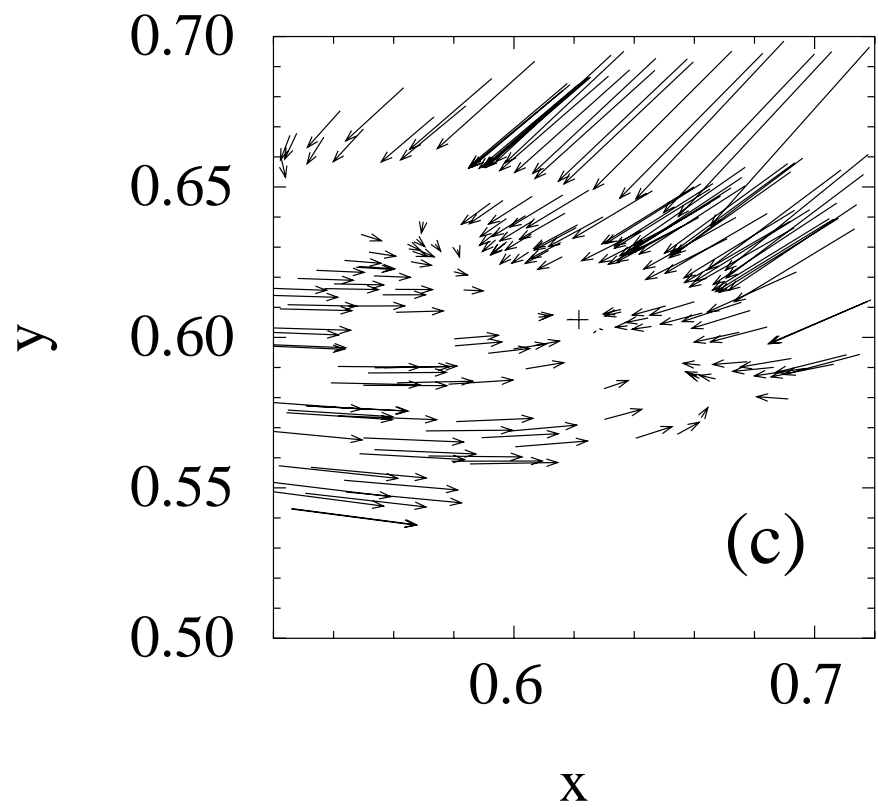
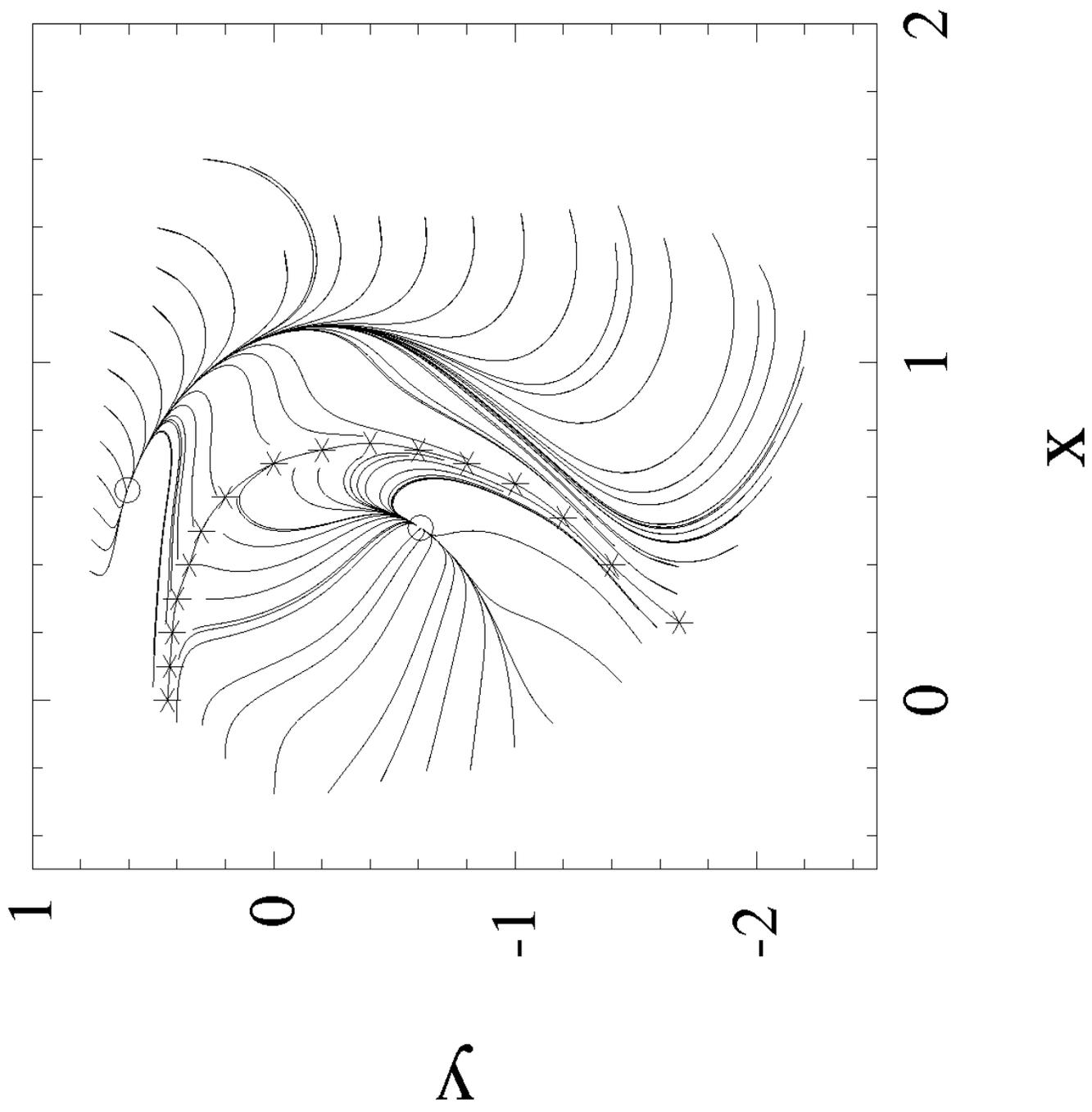
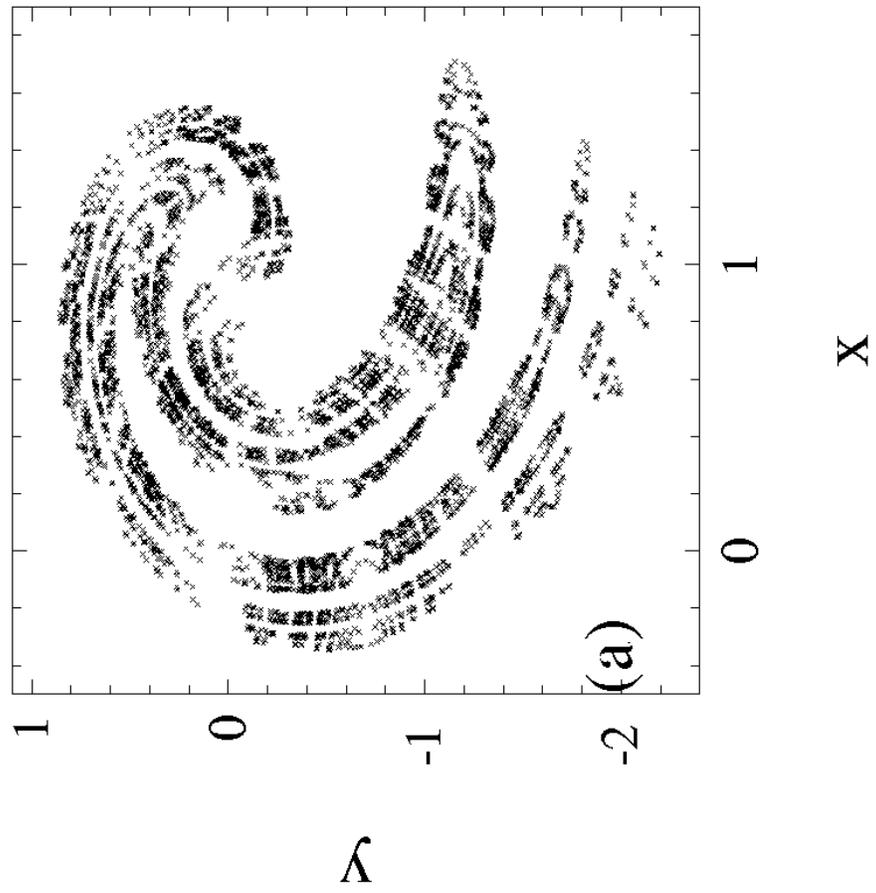
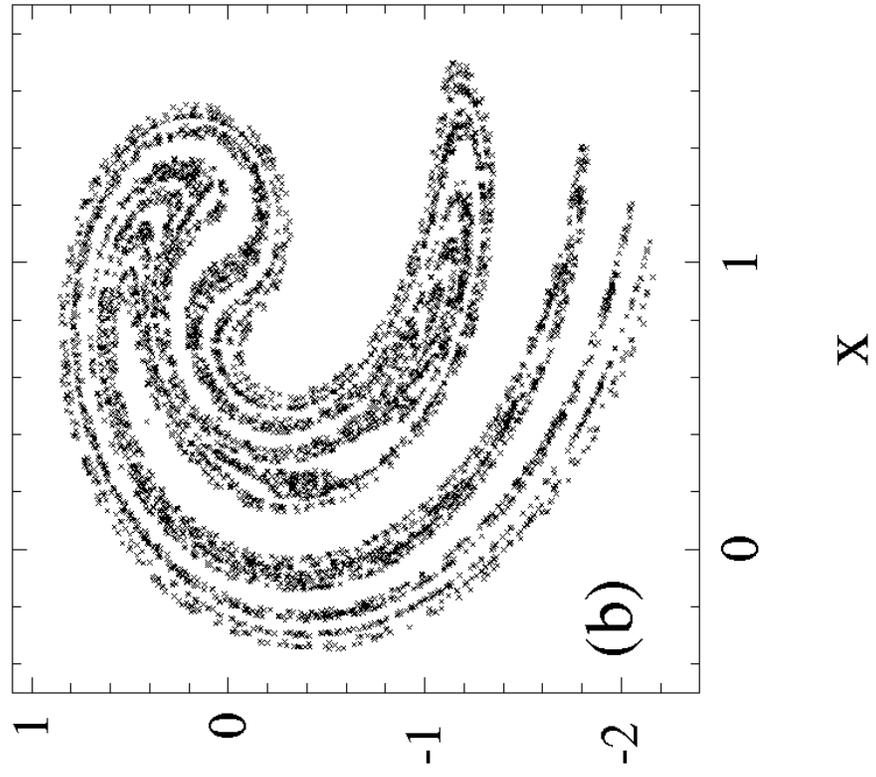
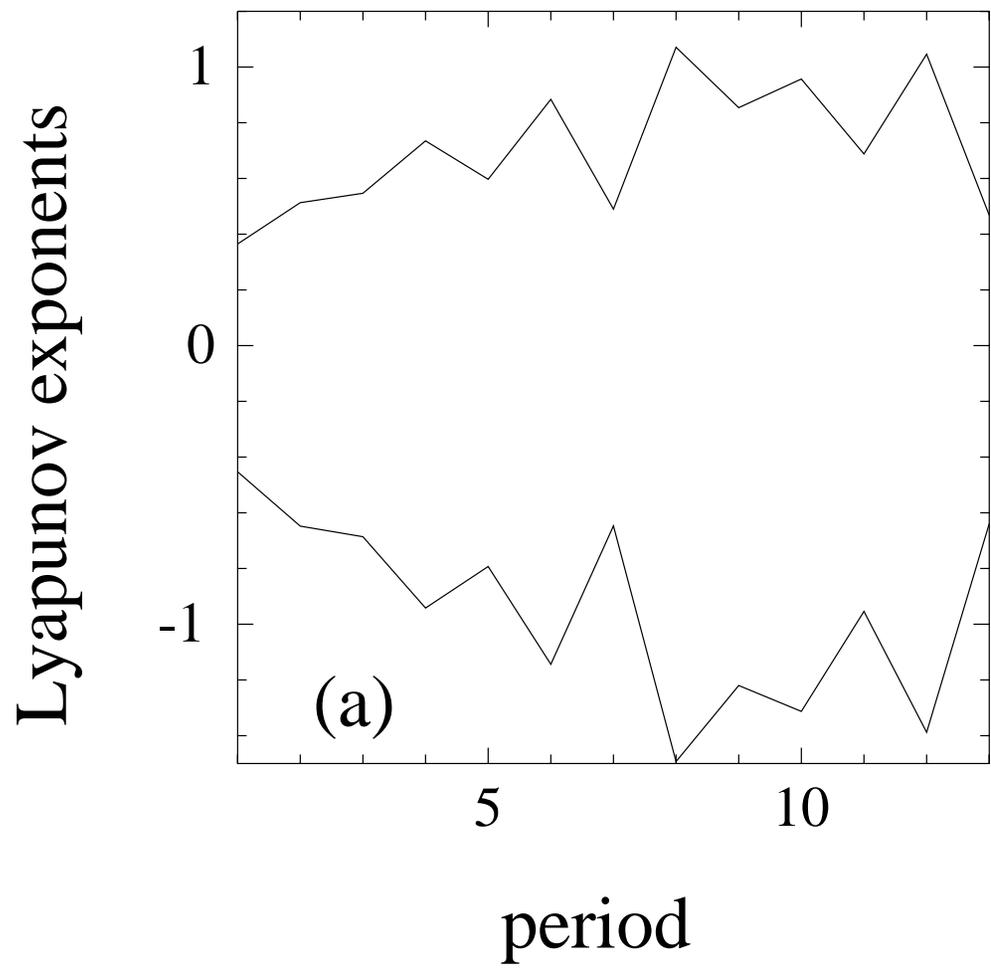
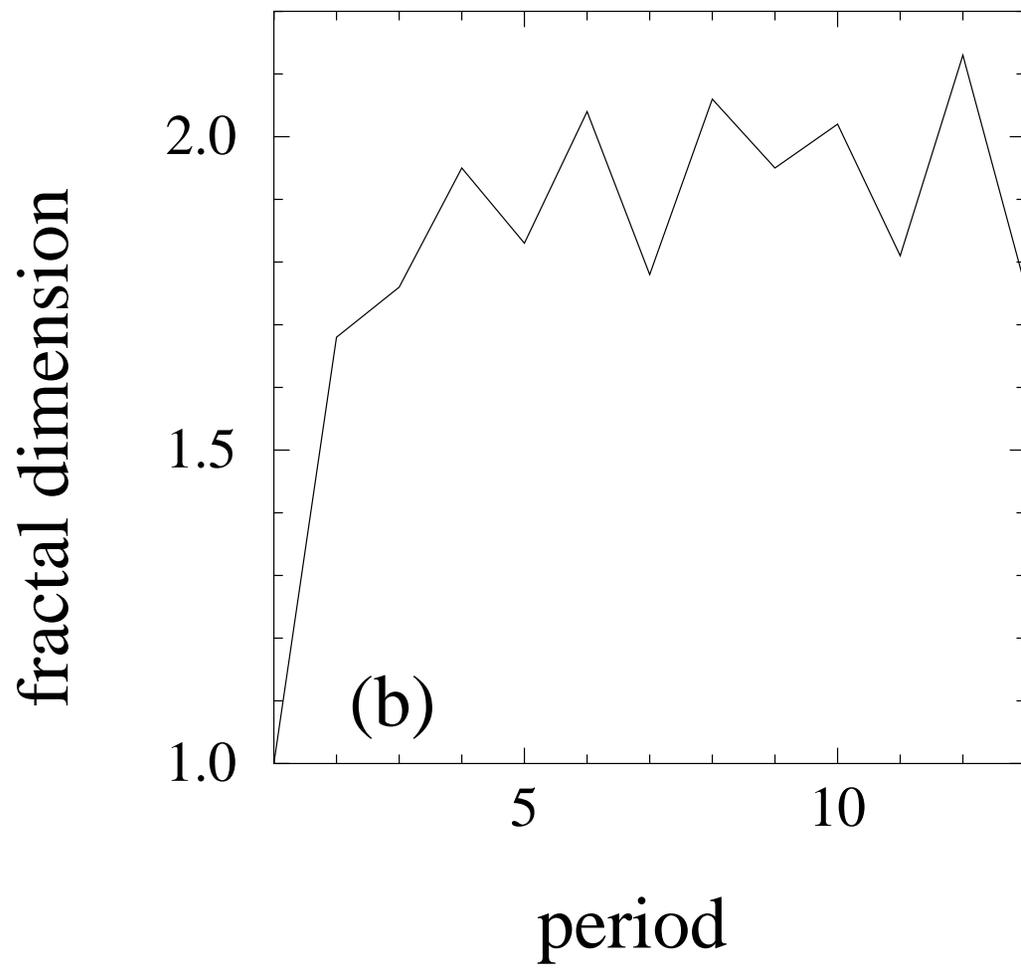
\end{document}